\documentclass[preprint,english,showpacs,floatfix]{revtex4}

\usepackage{babel,amsmath,amssymb,dcolumn,graphicx}

\begin{document}

\title{Perturbations of black $p$-branes}

\author{Elcio Abdalla\footnote{Electronic address: eabdalla@fma.if.usp.br}}
\affiliation{Instituto de F\'{\i}sica, Universidade de S\~ao Paulo,
  CP 66318, 05315-970, S\~ao Paulo-SP, Brazil}
\author{Owen Pavel Fernandez Piedra\footnote{Electronic address: opavel@ucf.edu.cu}}
\affiliation{ Departamento de F\'{i}sica y Qu\'{i}mica, Facultad de Mec\'{a}nica, Universidad de Cienfuegos, Carretera a Rodas,km 4, Cuatro Caminos, Cienfuegos, Cuba}
\author{Jeferson de Oliveira\footnote{Electronic address: jeferson@fma.if.usp.br}}
\affiliation{Instituto de F\'{\i}sica, Universidade de S\~ao Paulo,CP 66318, 05315-970, S\~ao Paulo-SP, Brazil}
\author{C. Molina\footnote{Electronic address: cmolina@usp.br}}
\affiliation{$^{3}$Escola de Artes, Ci\^{e}ncias e Humanidades, Universidade de
  S\~{a}o Paulo\\ Av. Arlindo Bettio 1000, CEP 03828-000, S\~{a}o
  Paulo-SP, Brazil}

\begin{abstract}
We consider black $p$-brane solutions of the low energy string action, computing scalar perturbations. Using standard methods, we derive the wave equations obeyed by the perturbations and treat them analytically and numerically. We have found that tensorial perturbations obtained via a gauge-invariant formalism leads to the same results as scalar perturbations. No instability has been found. Asymptotically, these solutions typically reduce to a $AdS_{(p+2)}\times S^{(8-p)}$ space, which, in the framework of Maldacena's conjecture, can be regarded as a gravitational dual to a conformal field theory defined in a  $(p+1)$-dimensional flat space-time. The results presented open the possibility of a better understanding the AdS/CFT correspondence, as originally formulated in terms of the relation among brane structures and gauge theories.
\end{abstract}



\maketitle
\section{Introduction}

String theory and the subsequent idea of branes have been, in recent years, the almost standard theory describing the physics of quantum space-time, especially near the Big Bang or even before it \cite{prebigbang}. The discovery of the relation between anti-de Sitter space physics and Conformal Field Theories on the boundary of that space, the so-called AdS/CFT correspondence \cite{adscft, largeN} implied further interest in the structure of the string-membrane theory. 

The $p$-brane extended solutions are considered fundamental in the understanding of the non-perturbative string theory regime. They interpolate $AdS_{p+2}\times S^{d-p-2}$ and $d$-dimensional Minkowski space-time \cite{gibbons}. This connection was important for the conjecture presented by Maldacena in 1997 \cite{adscft}, which opened the way for the gravitation-field theory dualities. In this context, a better understanding of the perturbative dynamics of the $p$-brane solutions are relevant for the structural aspects of the AdS/CFT correspondence and its latter extensions. Such extensions can provide new hints about Yang-Mills theory with special interest in what concerns the difficult question of a Quark Gluon Plasma, see for instance \cite{qgplasma, starinets, starinets2}. Besides, in the framework of AdS/CFT correspondence it is possible to study the glueball mass spectrum analyzing the dynamics of a scalar field in the near horizon limit of the black $p$-brane solutions \cite{csaki,boschi_filho1,alex_boschi}. The poles of the retarded function of the simplest glueball state, generated by the operator $\mathcal{O}=Tr(\it{F}^{2})$, are the quasinormal modes of the dual AdS black hole in the corresponding near horizon limit.

A fundamental feature of the $p$-brane backgrounds is the possible existence of event horizons. In this sense, they may be viewed as generalizations of the usual four-dimensional black holes. Perturbations of black hole solutions are well known \cite{chandra,ik} and several numerical methods exist, being under full control to handle the information gathered from such perturbations \cite{kokschprice,carlemoskono,abdwangetc}.

We intend here to first define a perturbation of a $p$-brane solution using standard separation of variables and subsequently treat, analytically and numerically, the wave equation for the scalar perturbation. The employed methods are largely independent, aiming to a cross-check of the results. We also consider gauge-invariant gravitational perturbations. The results turn out to
be exactly the same as scalar case.

One very recent work complement our analysis presented here \cite{vilson}. But although the presented work and \cite{vilson} are complementary and relevant in terms of the AdS/CFT correspondence, they treat different geometries and focus on different issues. The results presented in this paper address directly the role of the brane structure (in the sense presented in \cite{gibbons,adscft}) on the gravitation-field theory duality, specifically searching for possible instabilities.

The paper is organized as follows. Sec. II provides reviews $p$-brane background considered in this work. In Sec. III, the perturbative dynamics is formulated and developed, followed by Sec. IV and V where the non extreme and extremal scenarios are specifically treated. In Sec. VII some final comments are presented.

\section{$p$-brane solutions}

Solutions of ten dimensional Supergravity describing the so-called $p$-branes are well known. Let us consider the bosonic sector of type II Supergravity in ten dimensions, given by \cite{horowitz_strominger,largeN}:
\begin{gather}
\label{acao}
S=\frac{1}{(2\pi)^{7}(l_{s})^{8}}
\int dx^{10}\sqrt{-g}
\left[e^{-2\phi}\left(R +4\left(\nabla\phi\right)^{2}\right)
-\frac{2}{(8-p)!}F^{2}_{p+2}\right],
\end{gather}
where $l_{s}$ is the string length, $g$ the determinant of the metric
tensor $g_{ab}$, $R$ the Ricci scalar, $\phi$ the dilaton field and
$F_{p+2}$ the field strength of the potential $A_{p+1}$.

The solution of Einstein's equations with $N$ electric charges and $p$ dimensions is obtained from
the Ansatz \cite{horowitz_strominger}
\begin{equation}\label{linha}
ds^2 = ds^{2}_{10-p} +e^{\alpha}\sum^{p}_{i=1}dy^{i}dy^{i}\quad,
\end{equation}
where $ds^{2}_{10-p}$ is the line element with lorentzian signature in
$(10-p)$ dimensions, $\alpha$ is a function of $x$, that is the bulk's radial coordinate, and the meaning of $N$ as a charge
arises from the Gauss Law. We can write a full solution as
\begin{gather}
\label{metrica}
ds^{2}=-A(x)dt^2 + B(x)\left[dr^2 + r^2 d\Omega^{2}_{p-1}\right]
+C(x)dx^2+x^2 D(x)d\Omega^2_{8-p}\quad,
\end{gather}
where $A(x)=\left(1-(a/x)^{7-p}\right)
\left(1-(b/x)^{7-p}\right)^{-1/2}$,  $
C(x)=\left(1-(b/x)^{7-p} \right)^{\alpha_{1}}
\left(1-(a/x)^{7-p}\right)^{-1} $, 
$
B(x)=\sqrt{1-(b/x)^{7-p}}$, $ D(x)=\left(1-(b/x)^{7-p}\right)^{\alpha_{2}}
$,
with $\alpha_{1}=-\frac{1}{2}-\frac{(5-p)}{(7-p)}$ and 
$\alpha_{2}=\frac{1}{2}-\frac{(5-p)}{(7-p)}$. The mass per unit volume is
$M= \frac{1}{(7-p)\kappa_{1}}\left[\left(8-p\right)a^{7-p}-b^{7-p}\right] $, the electric charge $N=\frac{1}{\kappa_2}\left[ab\right]^{(7-p)/2}$,
$\kappa_{1}=(2\pi)^{7}d_{p}l_{p}^{8}$, $\kappa_{2}=d_{p}g_{s}l_{s}^{7-p}$, and
$g_{s}$ is the string coupling, $l_{p}$ the Planck length in ten
dimensions and $d_{p}=2^{5-p}\pi^{(5-p)/2}\Gamma((7-p)/2)$. Absence of naked singularities implies
\begin{equation}\label{massa_carga}
M\geq\frac{N}{(2\pi)^{p}g_{s}l_{s}^{p+1}}\quad .
\end{equation}

Considering the non-extreme scenario, the maximal extension of the metric describes a black brane geometry, with an event horizon located at $x=a$. If $p \ne 3$, a curvature singularity is present at $x=b$, while if $p=3$ we observe that, in addition to the outer horizon at $x=a$, there is also an inner horizon at $x=b$, with the singularity at $x=0$. That behaviour is observed in the  the Kretschmann scalar  $\mathcal{K}_{p}(x)=R_{abcd}R^{abcd}$, where $R_{abcd}$ are the components of Riemann tensor, as seen in the expression for the divergent term,
\begin{equation}
\mathcal{K}_{p}(x) \sim \frac{1}{\left(1-\left(\frac{b}{x}\right)^{7-p}\right)^{\delta(p)} x^{2(9-p)}}\quad,
\end{equation}
where
$ \delta(p) = \frac{1}{7-p}\left[(1+p)+2(5-p)\right] $ if $p$ is even, and 
$ \delta(p) = \frac{30p}{40}(p-1)(p-3)-\frac{p}{6}(p-1)(p-5) \\ +\frac{8p}{35}(p-3)(p-5)$ if $p$ is odd. 

For extremal $p$-branes the metric reads
\begin{gather}
\label{metrica_extrema}
ds^2=E(x)\left[-dt^2+dr^2 +r^2d\Omega_{p-1}^2\right] 
+ F(x)dx^2 + x^2G(x)d\Omega^{2}_{8-p}\quad ,
\end{gather}
where $E(x)=\sqrt{1-(a/x)^{7-p}}$, $F(x)=\left(1-(a/x)^{7-p}\right)^{\gamma_1}$, $G(x)=\left(1-(a/x)^{7-p}\right)^{\alpha_2}$, $
\gamma_1=\alpha_{1}-1$.

In the extreme case, the curvature singularity is located at $r=a$ and the metric does not have an extension if $p \ne 3$. We have a curvature singularity, but its structure depends on the value of $p$. If $p=6$ the singularity is time-like, and the proper definition of a Cauchy problem is delicate. On the other hand, if $p=0,1,2,5$, the singularity ($r=a$) is null \cite{largeN}, and therefore much milder. In spite of the absence of a event horizon, the manifold is globally hyperbolic, and the wave problem is well-posed. For the extreme case and $p=3$, there an analytic continuation of the metric beyond $r=a$ and we have again a black hole solution as pointed out in \cite{largeN}.

\section{Scalar and gravitational perturbative dynamics} 

We initially consider a massless scalar field in the background of our 10-dimensional solution.  We will show in the following that this scenario is more general. This perturbation is described by  the Klein-Gordon equation
\begin{equation}
\label{kg1operador}
\Delta_{10}\Phi\equiv\left\lbrack
\Delta_{p}(r,\theta_{(p-1)})
+\Delta_{10-p}(t,x,\lambda_{(8-p)})\right\rbrack\Phi =0 \, ,
\end{equation}
where the first term refers to the subspace $dr^2 + r^2
d\Omega^{2}_{p-1}$ and the second to the bulk coordinates
$(t,x,\lambda_{(8-p)})$. We denote the angular
coordinates in $d\Omega^{2}_{p-1}$ and $d\Omega^{2}_{8-p}$ respectively by  by  $\theta_{(p-1)}$ and $\lambda_{8-p}$.

Such equation can be separated by the Ansatz $\Phi(x^{A})=\sum_{l,m}R_{l}(r)Y_{lm}(\theta_{i})\sum_{L,q}\Psi_{L}(t,x)Y_{Lq}(\lambda_{j})$, where $Y_{lm}(\theta_{i})$ and $Y_{Lq}(\lambda_{j})$ are the well
known spherical harmonics in $(p-1)$ and $(8-p)$ dimensions
respectively \cite{eder}, resulting in the differential equations
\begin{equation}
\label{eq_brana}
\frac{1}{r^{(p-1)}}\frac{d}{dr}\left(r^{(p-1)}\frac{dR_{l}}{dr}\right) 
+ \left[\beta^2
  -\frac{l(l+p-2)}{r^2}\right]R_{l}=0,\\
\end{equation}
\begin{gather}
\label{eq_bulk}
-\frac{\partial^{2}\Psi_{L}}{\partial t^{2}}
+\frac{1}{A(x)}\Delta_{x}\Psi_{L}+u(x)\Psi_{L}=0 \, .
\end{gather}
where $u(x)= -\frac{A(x)}{B(x)}\left[\beta^{2} +
  \frac{B(x)}{x^{2}D(x)}L(L+7-p)\right]$. Moreover, $\beta$ is a constant arising from the brane \{$r$, $\theta_{(p-1)}$\} and bulk \{$t$, $x$, $\lambda_{(8-p)}$\} variables separation. Moreover, $\Delta_{x}$ is a differential operator given by
\begin{equation}
\Delta_{x}=\frac{
\frac{\partial}{\partial x}\left(\sqrt{A(x)B(x)C(x)D^{(8-p)}}x^{8-p}\frac{\partial}{\partial x}\right)
}{\sqrt{A(x)B(x)C(x)D^{(8-p)}}x^{8-p}} \, .
\end{equation}

The solution of equation (\ref{eq_brana}) is $R_{l}(r)=A_{1}r^{1-p/2} J_{\gamma}(\beta r) +A_{2}r^{1-p/2}Y_{\gamma}(\beta r)$,  with $\gamma=\frac{1}{2}\sqrt{p^2 - 4p +4 +4l(l+p-2)}$, $A_1$ and $A_2$
being constants, $J_{\gamma}(\beta r)$ and $Y_{\gamma}(\beta r)$ the Bessel functions. Finiteness at origin implies $A_{2}=0$ and  $R_{l}(r)=A_{1}r^{1-p/2}J_{\gamma}(\beta r)$. Therefore, $\beta$ has a continuous spectrum of allowed values, and we notice in (\ref{eq_bulk}) that the its square acts as a mass for the
Klein-Gordon field. Performing the same analysis for a time independent scalar field in the near horizon limit of the metric (\ref{metrica}), the $\beta^2$ parameter can be interpreted as the glueball mass.

A ``time independent approach'' can be explored expanding the function $\Psi_{L}(t,x)$ with a Laplace-like transform \cite{Nollert}. Within this approach, we obtain the equation
\begin{equation}\label{onda_bulk3}
\frac{d^{2}}{dr_{*}^2}Z_{L} + \left[k^2 - V(x)\right]Z_{L}=0\quad ,
\end{equation}
where we defined the tortoise coordinate as ${dr_{*}}/{dx}=\sqrt{C(x)/A(x)}$, 
$\Psi_{L}(t,x)=\int e^{i\omega t}b(x)Z_{L}(x) d\omega $ with
$b(x)=\frac{1}{x^{(8-p)/2}B(x)^{p/4}D(x)^{(8-p)/4}}$, $k^{2}=\omega^2 - \beta^2$ and  the  effective potential is given by the expression
\begin{gather}
V(x)=\left[\frac{A(x)}{B(x)}-1\right]\beta^{2} + \frac{A(x)}{x^{2}D(x)}L(L+7-p)
-\frac{1}{b(x)}\left[h(x)b(x)^{''}-g(x)b(x)^{'}\right]
\label{potencial_explicito}
\end{gather}
where the primes denotes differentiation with respect to $x$, $h(x)=A(x)/C(x)$, and  $g(x)=\frac{A(x)}{C(x)}\frac{d}{dx}\left\{\ln{\left[\frac{A(x)B(x)(D(x)x)^{8-p}}{C(x)}\right]}\right\}$.

We can also consider the problem of the linear perturbations using the
gauge-invariant formalism proposed by Ishibashi {\it{et al}}
\cite{ik}. In this formalism we expand the gravitational perturbations in terms
of tensor harmonics $\Pi_{ij}$, and perturbations of Einstein
equations are expressed as a group of equations for gauge invariant
quantities. Such quantities are grouped in three types: tensor,
vector and scalar. For the sake of simplicity, we only
consider in the following the tensor sector of gravitational
perturbations.
The spacetime is considered as describing an \(m+n\)-dimensional
manifold \(\mathcal{M}\), which is locally written as the warped
product $g_{\alpha\beta}dz^{\alpha}dz^{\beta}=g_{ab}(y)dy^{a}dy^{b}+f(y)\gamma_{ij}dx^{i}dx^{j}$,  where $\gamma_{ij}(y)$ is the metric of an \(n\)-dimensional maximally symmetric space of constant spatial curvature, and $g_{ab}(y)$ the metric of an arbitrary \(m\)-dimensional space time.

Following reference \cite{ik} the following equation for the
gauge-invariant quantity $H_{T}$ can be obtained:
\begin{equation}\label{ishi}
\Box H_{T} +\frac{8-p}{f}Dr\cdot DH_{T}-\frac{l(l+7-p)}{f^2}H_{T}=0,
\end{equation}
where $\Box$ is the D'Alembert operator written on the metric
$g_{ab}(y)$. Introducing in the above equation the master variable
\(\Phi=f^\frac{{8-p}}{2}H_{T}\) we found the same result that we
have already obtained from the scalar Klein-Gordon equation.

At this point it is appropriate to make the following important observation: the spectrum of
quasinormal frequencies for the scalar field perturbations contains
extra modes with respect to the tensor perturbations, because the
modes for the last case only appears for multipole numbers equal or
greater than 2. Thus, for the black $p$-brane, we need only to
consider a test scalar field perturbation. Extracting the \(l\geq2\)
terms for the obtained spectrum of scalar quasinormal frequencies,
we obtain the spectrum for the tensor gravitational perturbations.

\section{Non extreme case}

The effective potentials derived above determine the perturbative dynamics. Of particular importance for this dynamics are the quasi normal modes. They are defined as solutions of the wave equations which satisfy the in-going and out-going boundary conditions. These modes are particularly relevant in the intermediate time behavior of the perturbation.

With arbitrary $L$, two different and independent numerical tools will be used in this work to calculate the quasinormal frequencies: a ``frequency domain'' approach based on a sixth order WKB technique\cite{wkb_6}, and a ``time domain'' method  based on a numerical characteristic integration scheme \cite{Pullin,Brady,Abdalla}. Both algorithms are well established.
\begin{figure}[htb!]
\begin{center}
\resizebox{0.49\columnwidth}{!}{\includegraphics*{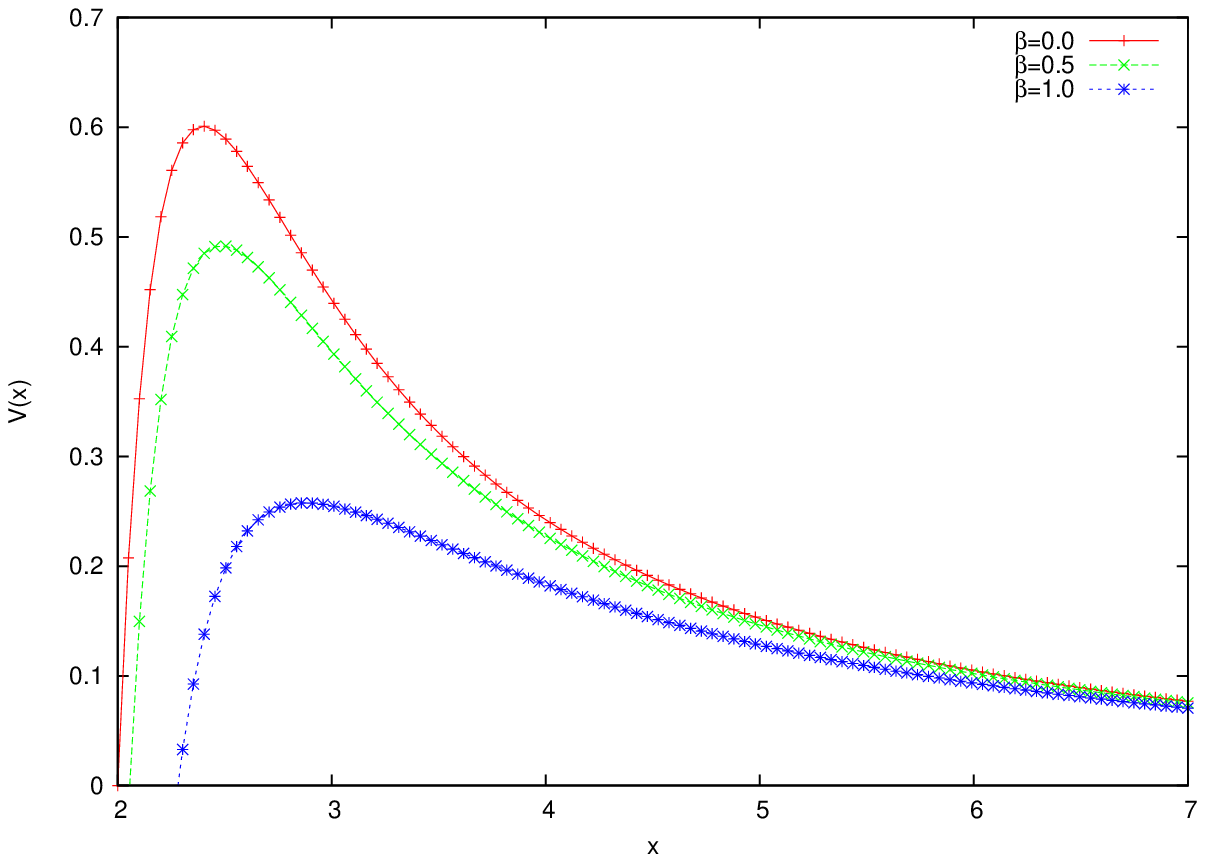}}
\resizebox{0.49\columnwidth}{!}{\includegraphics*{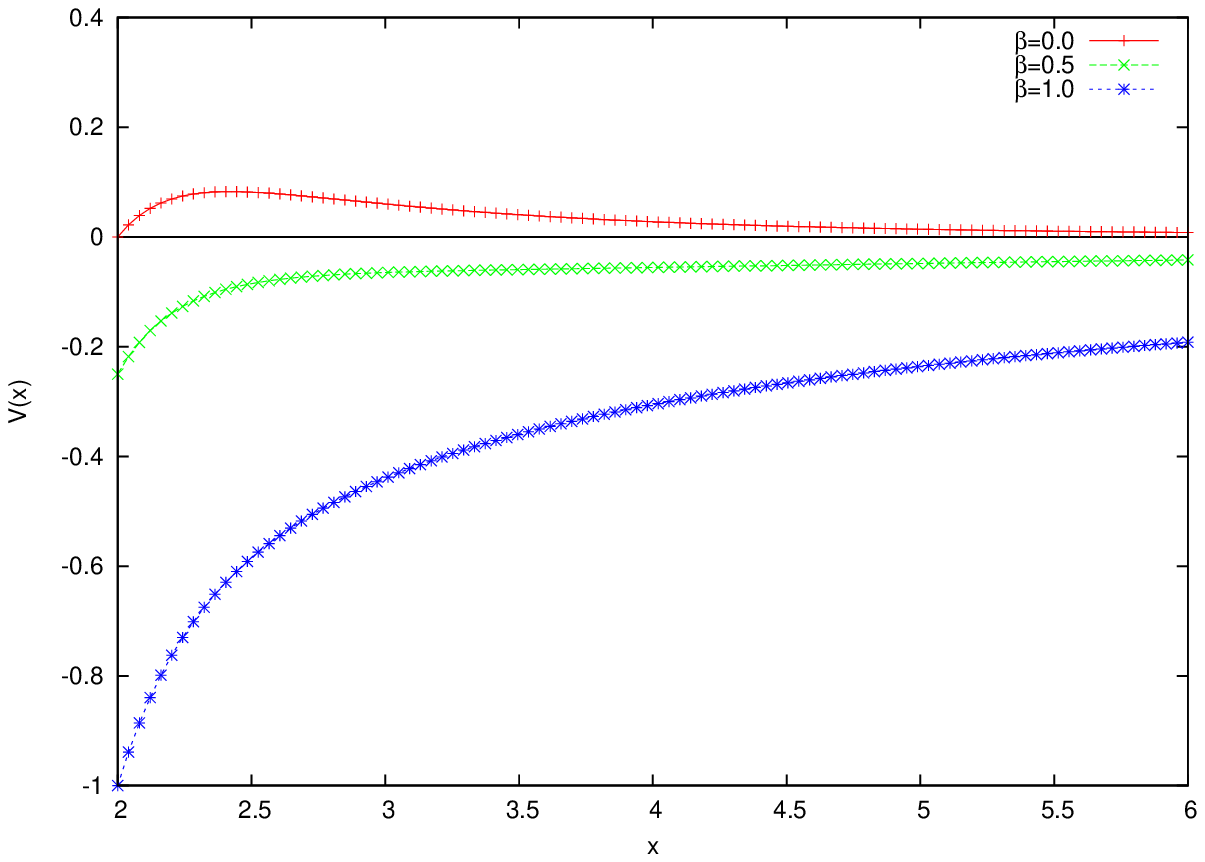}}
\caption{Effective potential for several values of $\beta$. The $p$-brane parameters of are $a=2$, $b=1$, $L=1$; and $p=3$ (left), $p=6$ (right).}
\label{potencial_beta_diff_zero}
\end{center}
\end{figure}

The WKB expressions are usually accurate and straitforward. But the approach is not generally applicable. For instance, in Fig. \ref{potencial_beta_diff_zero} the effective potencial is presented for a few values of $\beta$ with $p= 3, 6$. We observe that the maximum of the effective potential decreases as $\beta$ increases for a given $p$. For a sufficiently large value of  $\beta$ the potential becomes negative. This behavior appears explicitly for $p=6$ with $\beta=1$. Therefore, we cannot obtain the quasinormal frequencies for all values of $p$ and $\beta$ using the WKB formula. The instability for effective potentials that exhibit a negative gap is not excluded \cite{roman3,roman4}. Direct time integration can be used for such scenarios. \emph{We have found no instabillities after an extensive exploration with $\beta^2 \ge 0$. }

\begin{figure}[htb]
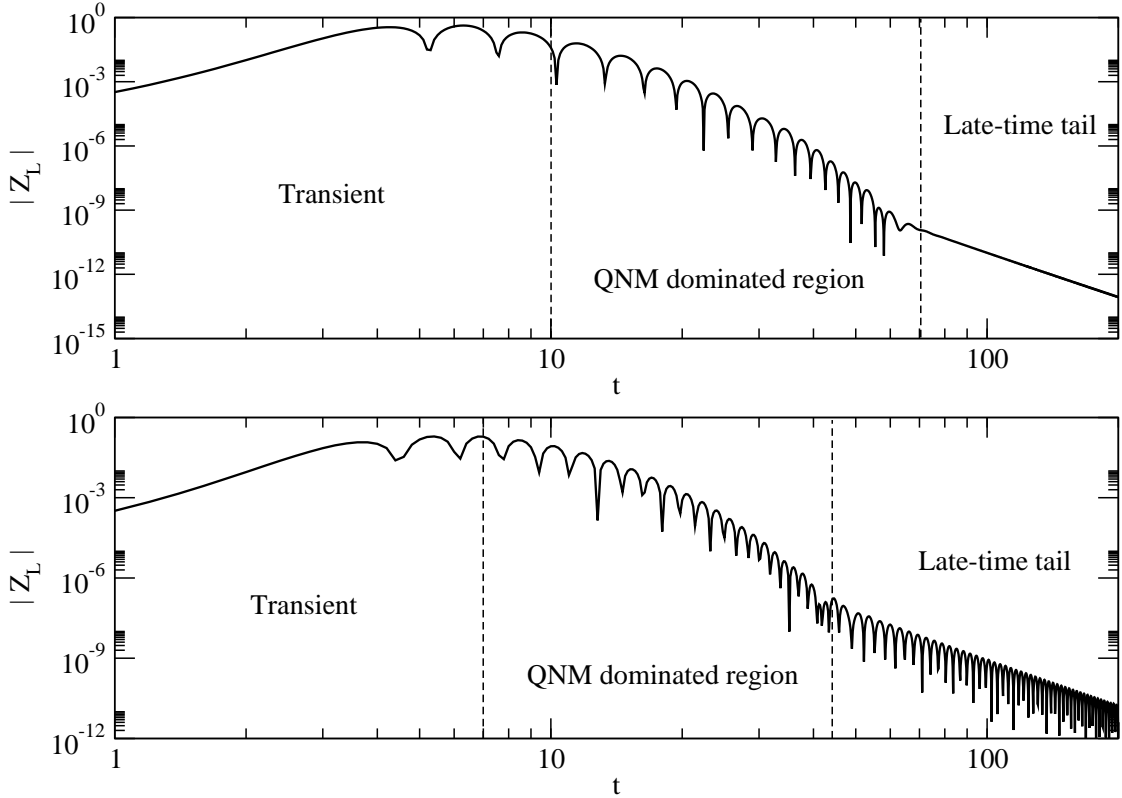

\resizebox{0.9\columnwidth}{!}{\includegraphics*{fig2a.eps}}\\
\resizebox{0.9\columnwidth}{!}{\includegraphics*{fig2b.eps}}
\caption{Log-log graph of the absolute value of $Z_{L}(t,x_{fixed})$. The quasinormal and tail phases are indicated. The $p$-brane parameters are $p=0$, $a=2$, $b=0.5$, $L=1$ and $\beta=0$ (top), $\beta=1$ (bottom).}
\label{profiles}
\end{figure}

Within the ``time-domain'' approach, we have observed the usual picture in the perturbative dynamics. After the initial transient regime, the quasinormal mode phase follows as well as a late-time tail. The tail phase is strongly dependent on the value of the parameter $\beta$. For $\beta=0$, we have a non-oscillatory power-law decay. But if $\beta\ne0$, the tail is oscillatory, with a power-law envelope. Typical profiles are shown in Fig. \ref{profiles}.
%

Given the potential, we use the sixth order WKB technique \cite{wkb_6} to obtain the quasinormal frequencies $k$. From the numerical data $Z_{L}(t,x_{fixed})$, it is possible to estimate the fundamental quasinormal frequency with reasonable accuracy. Some results from both methods are given in Tables \ref{frequencies1}, \ref{frequencies2} and \ref{high_overtone} for $\beta=0$. The concordance between them is good. However, notice that for $p=6$ and $L=0$, our result should be taken with reservation. Higher overtones are not accessible by the ``time-domain'' technique. The corresponding WKB results are presented in Table \ref{high_overtone}.


\begin{table}[htb!]\footnotesize
\caption{Fundamental quasinormal frequencies with $a=2$ and $b=0.5$ for $p=0,1,2,3$.}  
 \begin{center}
   \begin{tabular}{|c|c|c|c|c|}
      \hline
      \multicolumn{5}{|c|}{$p=0$}\\
      \hline
      \hline
      \multicolumn{3}{|c|}{WKB}&\multicolumn{2}{|c|}{Time evolution}\\
\hline
      $L$  & $\textrm{Re}(k)$ & $-\textrm{Im}(k)$  & $\textrm{Re}(k)$ & $-\textrm{Im}(k)$ \\
      \hline
0 & 1.2889  & 0.5506  & 1.250 (3.0) & 0.4980 (9.6) \\
\hline
1 & 1.5047  & 0.5876 & 1.606 (6.7) & 0.4867 (17.2)  \\
\hline
2 & 1.9638  & 0.4812 & 1.962 (0.092) & 0.4805 (0.15) \\
     \hline
    \hline
      \multicolumn{5}{|c|}{$p=1$}  \\
      \hline
      \hline
      \multicolumn{3}{|c|}{WKB}&\multicolumn{2}{|c|}{Time evolution}\\
      \hline
     $L$  & $\textrm{Re}(k)$ & $-\textrm{Im}(k)$  & $\textrm{Re}(k)$ & $-\textrm{Im}(k)$ \\
     \hline
 0 & 1.0812  & 0.4670 & 1.042 (3.6) & 0.4498 (3.7)    \\
\hline
1 & 1.3245  & 0.4963  & 1.604 (21.1) & 0.463 (6.7)    \\
\hline
2 & 1.7264  & 0.4301 & 1.725 (0.079) & 0.4295 (0.13)    \\
     \hline
    \hline
    \multicolumn{5}{|c|}{$p=2$}  \\
      \hline
      \hline
      \multicolumn{3}{|c|}{WKB}&\multicolumn{2}{|c|}{Time evolution}\\
      \hline
      $L$  & $\textrm{Re}(k)$ & $-\textrm{Im}(k)$  & $\textrm{Re}(k)$ & $-\textrm{Im}(k)$ \\
      \hline
 0 & 0.8714 & 0.3911 & 0.8346 (4.2) & 0.3926 (0.38)     \\
\hline
1 & 1.1311  & 0.4137 & 1.161 (2.64) & 0.3803 (8.1)     \\
\hline
2 & 1.488  & 0.3754 & 1.488 (0.013) & 0.3749 (0.13)     \\
     \hline
    \hline
\multicolumn{5}{|c|}{$p=3$}  \\
      \hline
      \hline
      \multicolumn{3}{|c|}{WKB}&\multicolumn{2}{|c|}{Time evolution}\\
      \hline
      $L$  & $\textrm{Re}(k)$ & $-\textrm{Im}(k)$  & $\textrm{Re}(k)$ & $-\textrm{Im}(k)$ \\
      \hline
  0 & 0.6633 & 0.3202   & 0.6376 (3.9) & 0.3279 (2.4)     \\
\hline
1 & 0.9284 & 0.3363 & 0.9413 (1.4) & 0.3204 (4.7)     \\
\hline
2 & 1.2489  & 0.3162 & 1.249 (0.0056) & 0.3157 (0.14)     \\
     \hline
 \end{tabular}
   \end{center}
\label{frequencies1}
  \end{table}


\begin{table}[htb!]\footnotesize
\caption{Fundamental quasinormal frequencies with $a=2$ and $b=0.5$ for $p=4,5,6$.}  
 \begin{center}
   \begin{tabular}{|c|c|c|c|c|}
 \hline
  \multicolumn{5}{|c|}{$p=4$}  \\
      \hline
      \hline
      \multicolumn{3}{|c|}{WKB}&\multicolumn{2}{|c|}{Time evolution}\\
      \hline
      $L$  & $\textrm{Re}(k)$ & $-\textrm{Im}(k)$  & $\textrm{Re}(k)$ & $-\textrm{Im}(k)$ \\
      \hline
  0 & 0.4632 & 0.2514 & 0.4449 (4.0) & 0.2555 (1.6)    \\
\hline
1 & 0.7211 & 0.2607  & 0.7244 (0.46) & 0.2438 (6.4)    \\
\hline
2 & 1.0081  & 0.2512 & 1.008 (0.012) & 0.2509 (0.13)     \\
     \hline
      \hline
\multicolumn{5}{|c|}{$p=5$}  \\
      \hline
      \hline
      \multicolumn{3}{|c|}{WKB}&\multicolumn{2}{|c|}{Time evolution}\\
      \hline
     $L$  & $\textrm{Re}(k)$ & $-\textrm{Im}(k)$  & $\textrm{Re}(k)$ & $-\textrm{Im}(k)$ \\
      \hline
 0 & 0.2825 & 0.1828 & 0.2697 (4.5) & 0.1990 (8.8)     \\
\hline
1 & 0.5179 & 0.1843 & 0.5187 (0.16) & 0.1828 (0.83)    \\
\hline
2 & 0.7690 & 0.1804 & 0.7691 (0.010) & 0.1802 (0.082)    \\
     \hline
    \hline
  \multicolumn{5}{|c|}{$p=6$}  \\
      \hline
      \hline
      \multicolumn{3}{|c|}{WKB}&\multicolumn{2}{|c|}{Time evolution}\\
      \hline
      $L$  & $\textrm{Re}(k)$ & $-\textrm{Im}(k)$  & $\textrm{Re}(k)$ & $-\textrm{Im}(k)$ \\
      \hline
 0 & 0.3135 & 0.05970& 0.1485 (52.6) & 0.1290 (116.1)   \\
\hline
1 & 0.3608 & 0.1154 &  0.3616 (0.22) & 0.1150 (0.34)    \\
\hline
2 & 0.5890 & 0.1135 &  0.5889 (0.021) & 0.1134 (0.042)   \\
     \hline
\end{tabular}
   \end{center}
\label{frequencies2}
  \end{table}

\begin{table}[htb!]\footnotesize
\caption{High overtone quasinormal frequencies with $a=2$ and $b=0.5$.}  
 \begin{center}
   \begin{tabular}{|c|c|c|c|c|c|}
      \hline
      \multicolumn{4}{|c|}{$p=0$}&\multicolumn{2}{|c|}{$p=1$}\\
      \hline
      \hline
      $L$ &$n$ &$\textrm{Re}(k)$ & $-\textrm{Im}(k)$ &$\textrm{Re}(k)$ & $-\textrm{Im}(k)$\\
      \hline
   1 & $1$ &0.985828   &1.79911  &0.892835 &1.58205 \\
\hline
     2 & $1$ &1.47092   &1.61706   &1.3581 &1.39048  \\
\hline
     2 & $2$ &0.408755    &2.80627  &0.538849  &2.55582  \\
     \hline
    \hline
 \multicolumn{4}{|c|}{$p=2$}&\multicolumn{2}{|c|}{$p=3$}\\
      \hline
      \hline
      $L$ &$n$ &$\textrm{Re}(k)$ & $-\textrm{Im}(k)$ &$\textrm{Re}(k)$ & $-\textrm{Im}(k)$\\
      \hline
  1 & $1$ & 0.798307 & 1.34085   & 0.693083 & 1.09224    \\
\hline
     2 & $1$ & 1.22235 &1.18843     & 1.0673  &  0.990437 \\
\hline
     2 & $2$ &0.638727  & 2.20914   &0.690423  & 1.82098  \\
     \hline
    \hline
 \multicolumn{4}{|c|}{$p=4$}&\multicolumn{2}{|c|}{$p=5$}\\
      \hline
      \hline
     $L$ &$n$ &$\textrm{Re}(k)$ & $-\textrm{Im}(k)$ &$\textrm{Re}(k)$ & $-\textrm{Im}(k)$\\
      \hline
  1 &  1 & 0.572698 & 0.841449    & 0.439874 & 0.587662\\
      \hline
 2 &  1 & 0.895481 &0.781983     & 0.710916  &  0.55703\\
      \hline
  2 &  2 &0.681196  & 1.40916   &0.609848  & 0.980543 \\
     \hline
 \end{tabular}
   \end{center}
\label{high_overtone}
  \end{table}

\begin{table}[htb!]\footnotesize
 \begin{center}
   \begin{tabular}{|c|c|c|c|}
\hline 
\multicolumn{4}{|c|}{$p=6$}\\
      \hline
      \hline
      $L$ &$n$ &$\textrm{Re}(k)$ & $-\textrm{Im}(k)$ \\
      \hline
   $ 1 $&$1$&  0.325148  & 0.365934   \\
      \hline
  $ 2 $& $1$ & 0.568875  &  0.345275 \\
      \hline
  $ 2 $&$2$ & 0.537292   & 0.587244 \\
     \hline
 \end{tabular}
\end{center}
\label{high_overtones6}
\end{table}

\begin{figure}[htb!]
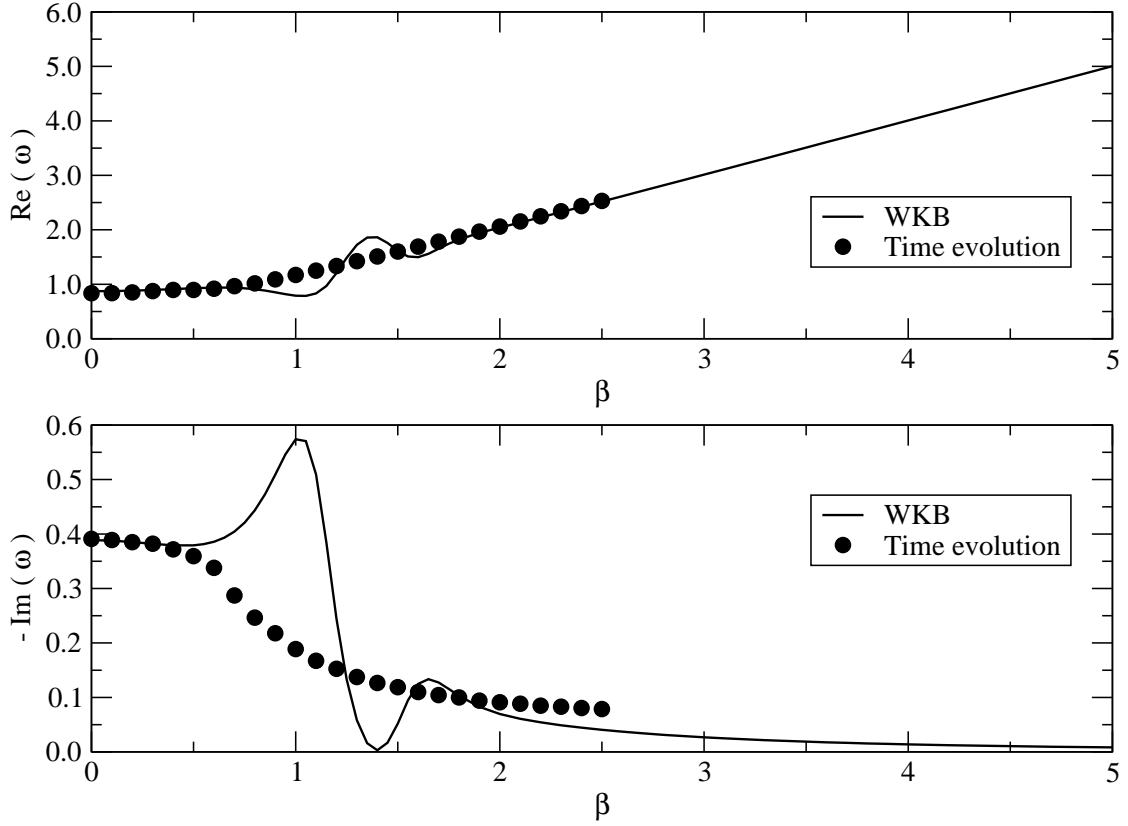

\begin{center}
\resizebox{0.9\columnwidth}{!}{\includegraphics*{fig3a.eps}}\\
\resizebox{0.9\columnwidth}{!}{\includegraphics*{fig3b.eps}}
\caption{Effect of $\beta$ on the behavior of $\omega$ for $p=2$ with $a=2$, $b=0.5$ and $L=0$. Two different numerical methods were employed. They are consistent for small and large enough $\beta$, but discrepant near $\beta=1$.}
\label{modos_beta_p=2}
\end{center}
\end{figure}

The dependence of the frequencies $\omega=\sqrt{k^2 + \beta^2}$ on $\beta$ was also investigated. Both WKB and direct integration methods were employed, although the time evolution approach is not applicable for large $\beta$, since in this regime the massive tail dominates from a very early time. Nevertheless, it should be reliable for small $\beta$. Generally, we observed that for large values of the mass parameter, as $\beta$ increases the frequencies becomes more oscillatory and less damped. One intriguing point was seen in a specific choice of parameters, namely $a=2$, $b=0.5$, $L=0$ and $p=2$. In this case, the WKB and time evolution methods give discrepant results near $\beta=1$, as shown in Fig. \ref{modos_beta_p=2}.

\begin{figure}[htb]
\begin{center}
\resizebox{0.49\columnwidth}{!}{\includegraphics*{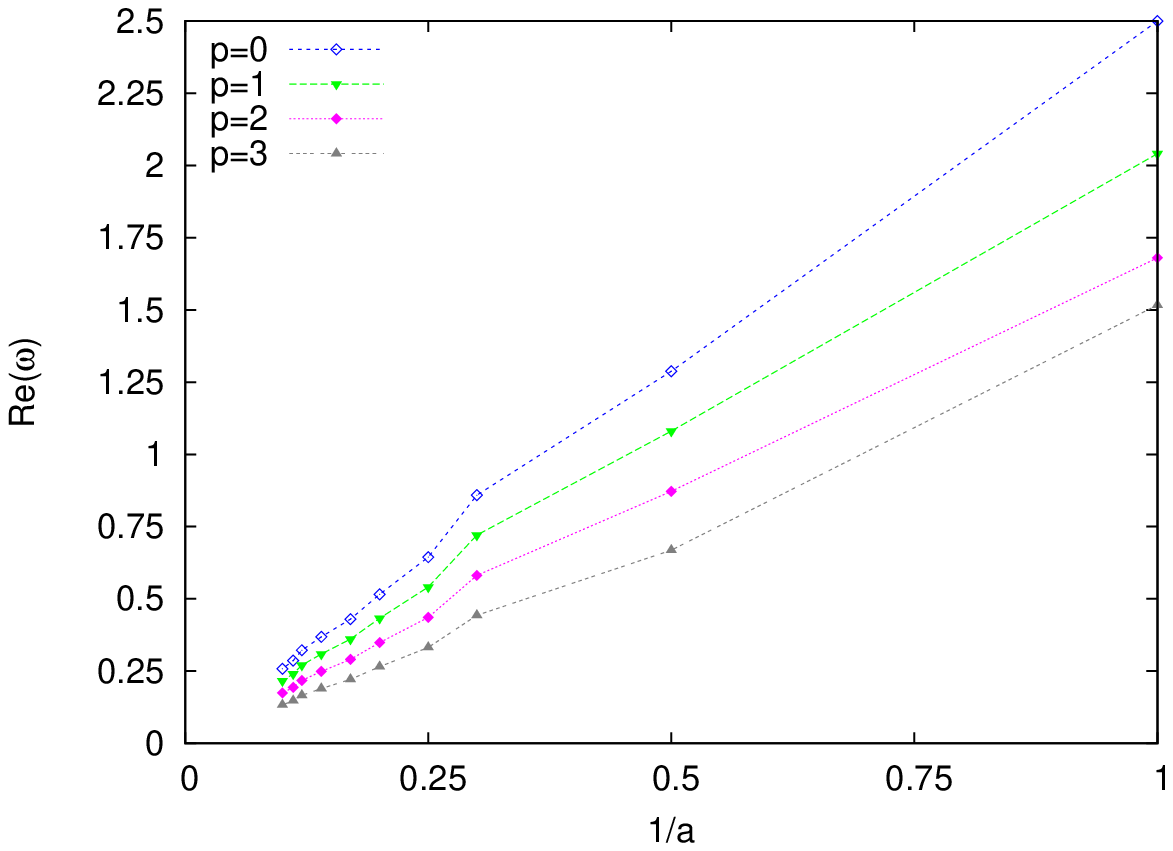}}
\resizebox{0.49\columnwidth}{!}{\includegraphics*{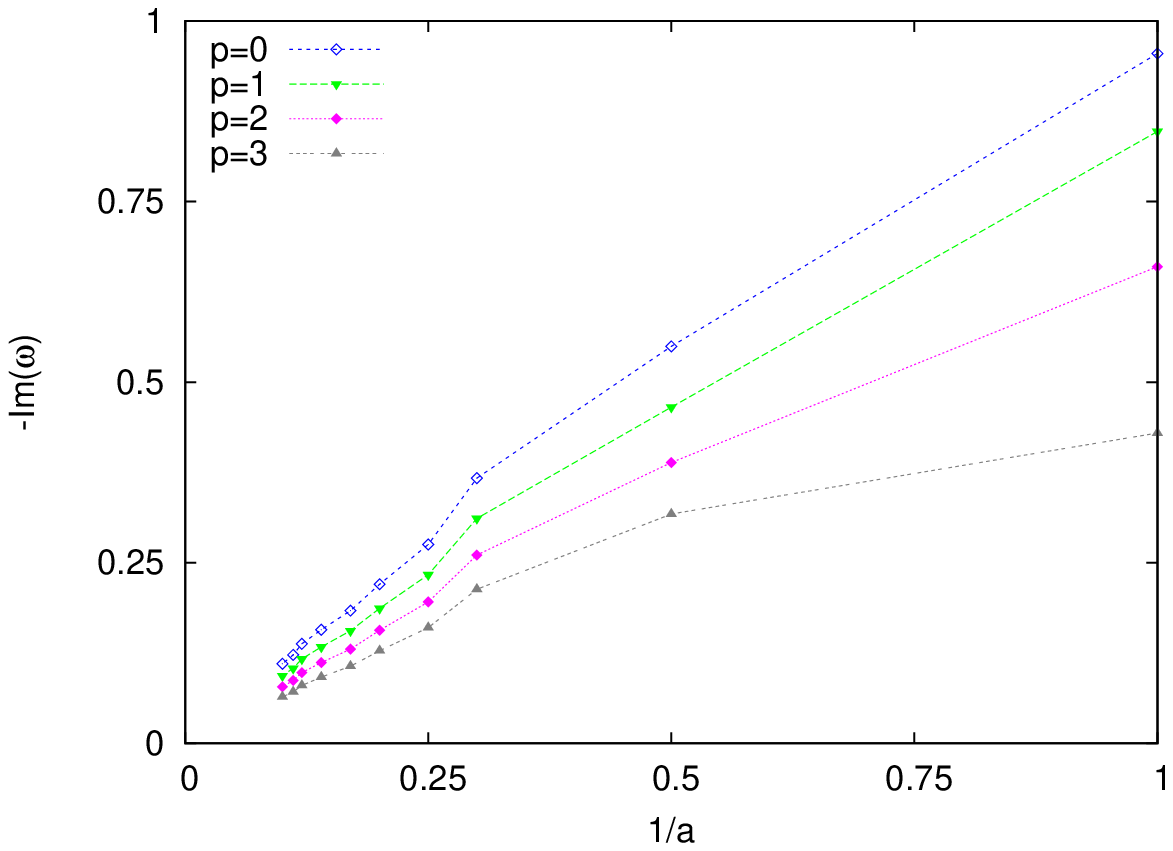}}
\caption{Effect of the $a$ parameter on quasinormal frequency. The $p$-brane parameters of are $b=0.5$, $L=0$; and $\beta=0$. }
\label{multiplot_p=0_var_a_1}
\end{center}
\end{figure}

It is worth noticing that the frequency $k$ show an almost scaling
behaviour on functions of $a^{-1}$, as shown in Fig. \ref{multiplot_p=0_var_a_1}.
That happens for the imaginary as well as for the real parts of $k$
except for very small values of $a$. We found a different behaviour
just in the case $L=2$, $n=2$, for the values of $a<2$ near the
extremal case $a=b$. No instability has been found. For higher
dimensions the real and imaginary parts of the frequency decreases.
An exception is the case $L=2$, $n=2$: the real part of the frequency
increases in the range $ 0\leq p \leq 3$  and decreases for the
others values of $p$, but the imaginary part decreases when
$p$ increases as for all others values of $L$ and $n$ that we
considered in this work. We have found that  for a given value of $L$  increasing the overtone number $n$ the frequencies become more damped, as we expected.

Although in general the calculation of the quasinormal frequencies can be only made using numerical methods, in the present scenario there is an important limit where an analytic expression is available. Expanding the effective potential in terms of small values of $1/L$ and using the WKB method in the lowest order (which is exact in this limit), we obtain:
\begin{equation}
\omega^{2}=L^{2}\Gamma (x_m) - i\left(n+\frac{1}{2}\right)L\Lambda (x_m) \, ,
\end{equation}
where  $\Gamma(x) = \frac{A(x)}{x^{2}D(x)}$, $\Lambda(x) =
-\frac{2A(x)}{C(x)}\sqrt{\frac{\Gamma(x)^{'}}{2}\left[\ln{(A(x)/C(x))}\right]^{'}+\Gamma^{''}}
$. The peak of effective potential is determined by $V(x)'$, and occurs at $x_{m}=\left[-2c_{1}/\left(c_{2}+(c_{2}^{2}-8c_{1})^{1/2}\right)\right]^{1/(7-p)}$, with $c_{1}=(7-p)(ab)^{7-p}$ and $c_{2}=-(9-p)a^{7-p}$.

Far from the horizon the effective potential (with $\beta=0$), in terms of $r_{\star}$, assumes the form
\begin{equation}
V(r_{\star}) = \begin{cases}
              \left(L + \frac{8-p}{2}\right) 
              \left(L + \frac{6-p}{2}\right)
              \frac{1}{r_{\star}^{2}} 
              + \mathcal{O} \left( \frac{1}{r_{\star}^{8-p}} \right) \\
              \textrm{if} \,\, 0 \le p < 6 \\
                                       & \\ 
                L (L + 1) \left[ \frac{1}{r_{\star}^{3}} 
                               + (2a - b)  \frac{\ln r_{\star}}{r_{\star}^{4}}
                \right]
              + \mathcal{O} \left( \frac{\ln r_{\star}}{r_{\star}^{5}} \right)\\ 
              \textrm{if} \,\, p = 6 
              \,\,\textrm{and}\,\, L = 0  \\
                                       & \\
              L (L + 1) \left[ \frac{1}{r_{\star}^{2}} 
                               + (2a - b)  \frac{\ln r_{\star}}{r_{\star}^{3}}
                \right]
              + \mathcal{O} \left( \frac{\ln r_{\star}}{r_{\star}^{4}} \right)\\ 
              \textrm{if} \,\, p = 6 
              \,\,\textrm{and}\,\, L > 0 &

              \end{cases}
\end{equation}
With this effective potential, it is shown \cite{Price,Ching} that an initial data with compact support evolves, at late time, according to
\begin{equation}
\Psi_{L} \sim t^{-\alpha(p,L)}\, .
\end{equation}
Therefore, at asymptotically late times the massless perturbation decay as a power-law tail. 

\begin{figure}[htb]
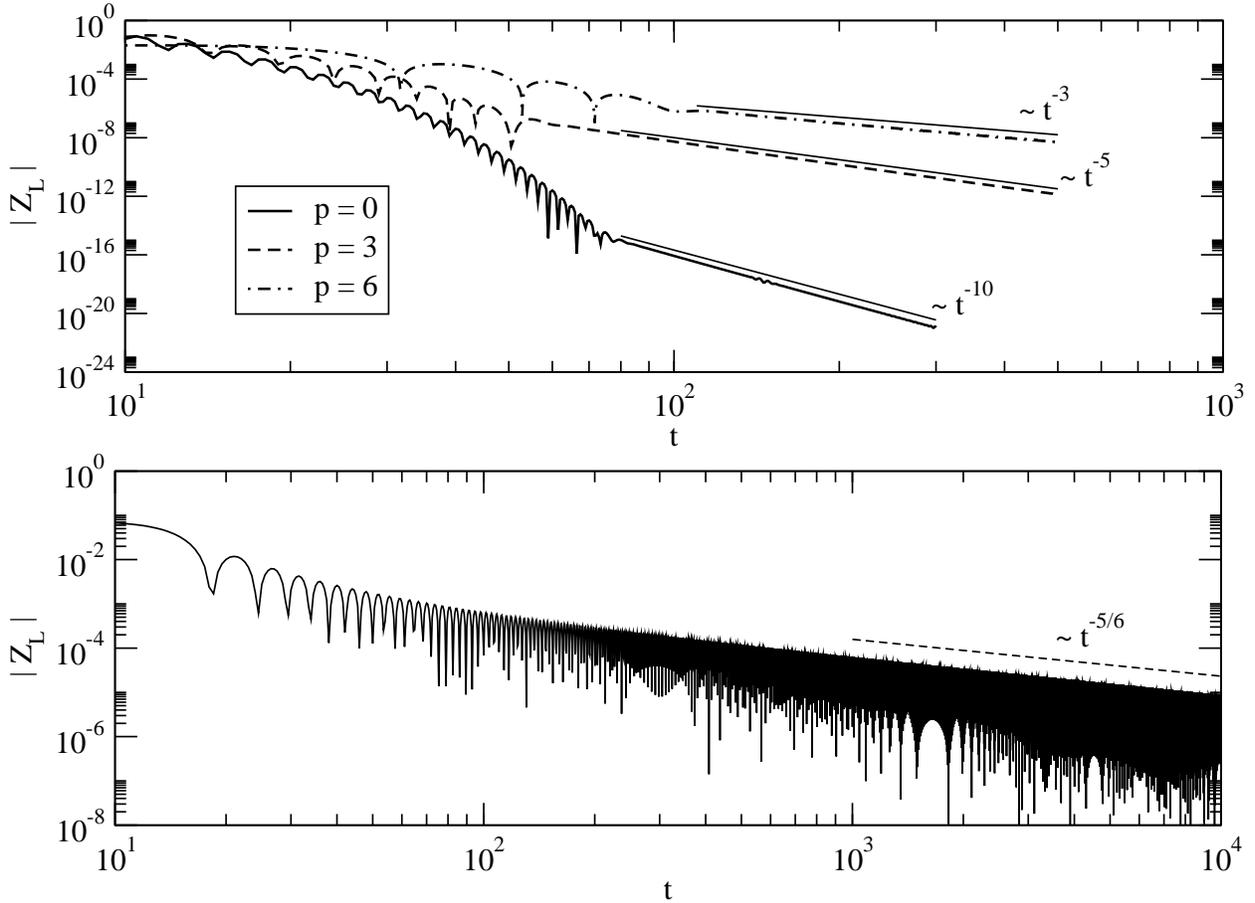

\resizebox{\columnwidth}{!}{\includegraphics*{fig5a.eps}}\\
\resizebox{\columnwidth}{!}{\includegraphics*{fig5b.eps}}
\caption{(Top) Tails for several values of $p$. The power-law  coefficients estimated from the numerical data (with $t>250$) are: -10.01 $(p=0)$, -5.07 $(p=3)$ and -3.15 $(p=6)$. The analytical results (indicated by straight lines) are: -10 $(p=0)$, -5 $(p=3)$ and -3 $(p=6)$. The $p$-brane parameters are $a=2$, $b=0.5$, $L=0$ and $\beta=0$. (Bottom) Massive tail for $p=6$. The envelope power-law coefficient estimated from the numerical data (with $t>9000$) is $-0.84$. The analytical result (indicated by a straight line) is $-5/6 \approx -0.833$. The $p$-brane parameters are $a=2$, $b=0.5$, $L=0$ and $\beta=1$.}
\label{tails}
\end{figure}

The power-law coefficient $\alpha(p,L)$ reflects the potencial asymptotic behavior. For $p=1,3,5,6$, $\alpha(p,L)$ can be analitycally determined using the results in \cite{Ching}:
\begin{equation}
\alpha (p,L) = \begin{cases}
              2L - p + 8 & 
              \,\,\textrm{with} \,\, p = 1,3,5  \\
              2L + 3 & 
              \,\,\textrm{with} \,\, p = 6  \\
              \end{cases}
\end{equation}
For $p=0,2,4$, our numerical resuts suggest a similar expression
\begin{equation}
\alpha (p,L) = 2L - p + 10 \,\,\textrm{with} \,\, p = 0,2,4  
\end{equation}
The tails are confirmed by the time-dependent approach. We illustrate these results in Fig. \ref{tails}.

In the massive case, the asymptotic form of the effective potential changes. For large $r_{\star}$ we have
\begin{equation}
V(r_{\star}) = \begin{cases}
              \beta^2 +
              \left(L + \frac{8-p}{2}\right) 
              \left(L + \frac{6-p}{2}\right)
              \frac{1}{r_{\star}^{2}} 
              + \mathcal{O} \left( \frac{1}{r_{\star}^{8-p}} \right) \\
              \textrm{if} \,\, 0 \le p < 5 \\
                                       & \\ 
              \beta^2 +
              \left[\beta^2 +
              L^2 + 2L + \frac{3}{4}
              \right] 
              \frac{1}{r_{\star}^{2}} 
              + \mathcal{O} \left( \frac{1}{r_{\star}^3} \right) \\
              \textrm{if} \,\, p = 5  \\
                                       & \\ 
              \beta^2 \left( 1 + \frac{b-a}{r_{\star}} \right)
              + \mathcal{O} \left( \frac{1}{r_{\star}^{3}} \right)\\ 
              \textrm{if} \,\, p = 6 
              \,\,\textrm{and}\,\, L = 0  \\
                                       & \\
              \beta^2 \left( 1 + \frac{b-a}{r_{\star}} \right)
              + \left[ \beta^2 b (b-a) \right. \\
                \left. + L(L+1) \right]\frac{1}{r_{\star}^2}
              + \mathcal{O} \left( \frac{\ln r_{\star}}{r_{\star}^{3}} \right)\\ 
              \textrm{if} \,\, p = 6 
              \,\,\textrm{and}\,\, L > 0 &

              \end{cases}
\end{equation}
We have observed from the numerical simulations that the late-time tail have the form
\begin{equation}
\Psi_L \sim \sin(\beta t) t^{-\gamma (p,L)} \, .
\end{equation}
If $p=6$, the results in \cite{koyama1,koyama2,roman5} apply, and the coefficient in the power-law envelope can be determined analytically: $\gamma (p=6,L) = 5/6$. This result is illustrated in Fig.\ref{tails}. For other values of $p$ the analytical problem remains open.

\section{Extreme case}

The analysis of the extreme case geometry is more subtle. If $p=3$, we have a black hole solution and the problem is clearly formulated. If $p=6$, we have a naked time-like singularity and the Cauchy problem is not well-posed (without additional conditions at the singularity). This class of solution will not be treated in the present work.

The novelty is the geometry with a null singularity. As discussed before, we have a well-posed initial value problem. We propose here to \emph{define} the quasi normal modes in the same way they were defined in the black hole scenario. This definition will be justified considering the wave problem in the following.

The effective potential for the scalar field perturbation in the extreme case scenario is obtained by taking $a=b$ in (\ref{potencial_explicito}). This potential  looks similar to the non extreme case analog, and in terms of the tortoise coordinate, it tends to zero as $r_{\star}\rightarrow -\infty$ and $r_{\star}\rightarrow \infty$, what implies that the effective one-dimensional wave problem is similar to the previous non-extreme case. A bounded perturbation will therefore decay in time, what justifies the quasi normal mode definition adopted. As a side remark, we observe that for $p=6$ the potential diverges near the horizon, a consequence of the time-like nature of the singularity at $r=a$.
We have computed the quasi normal frequencies for $p<5$. The results are shown in Table \ref{frequencies_extremo}. We have sensible differences, by factors of order three.

For $L=2$, from $n=0$ to $n=1$
we observe an increase in the decay rate.
We found  that the imaginary part  increases, in the case $p=0$ from $L=2$,
$n=2$ to $n=1$,  in contrast with the behaviour found in the non
extreme case. Otherwise,
results are very similar to the non extreme case.

\begin{table}[htb!]\footnotesize
\caption{Scalar quasinormal frequencies for the extreme case $(a=b=1)$.}  
 \begin{center}
   \begin{tabular}{|c|c|c|c|c|c|}
      \hline
      \multicolumn{4}{|c|}{$p=0$}&\multicolumn{2}{|c|}{$p=1$}\\
      \hline
      \hline
      $L$ &$n$ &$\textrm{Re}(k)$ & $-\textrm{Im}(k)$ &$\textrm{Re}(k)$ & $-\textrm{Im}(k)$\\
      \hline
   0  & 0 &  2.49971  & 0.955112   & 2.04071  & 0.847233  \\
\hline
   1 & 0 & 3.07066  &  1.01771   &  2.68902  &  0.863596 \\
\hline
  1 & 1 & 2.41319   &  2.08326   & 1.9794 & 2.17509 \\
\hline
  2 & 0 &  3.88648  & 0.931587   & 3.47584 &0.802949  \\
\hline
     2 & 1 &  3.15979  & 2.7734  & 2.88255 &2.40614  \\
\hline
 2 & 2 & 0.0876808   &  2.4072       &0.978905  &3.54121  \\
\hline
\hline
 \multicolumn{4}{|c|}{$p=2$}&\multicolumn{2}{|c|}{$p=3$}\\
      \hline
      \hline
      $L$ &$n$ &$\textrm{Re}(k)$ & $-\textrm{Im}(k)$ &$\textrm{Re}(k)$ & $-\textrm{Im}(k)$\\
      \hline
  $ 0 $ & $0$& 1.6804 & 0.659786 & 1.51662 & 0.429531 \\
      \hline
  $ 1 $& $0$ & 2.35616 & 0.685455 & 2.09221 &0.522615 \\
      \hline
 $ 1 $& $1$ & 1.79834 & 1.90349  & 1.7917 & 1.46652 \\
      \hline
  $ 2 $& $0$ & 3.10546 & 0.658724 & 2.79837 &  0.513388 \\
     \hline
 $ 2 $& $1$ & 2.67848 &1.97719  & 2.52534  &  1.55801 \\
    \hline
 $ 2 $& $2$ &1.52044  & 3.22954   &1.92728  & 2.57206 \\
     \hline
 \end{tabular}
   \end{center}
\label{frequencies_extremo}
  \end{table}

\begin{table}[htb!]\footnotesize
 \begin{center}

   \begin{tabular}{|c|c|c|c|}
\hline 
\multicolumn{4}{|c|}{$p=4$}\\
      \hline
      \hline
      $L$ &$n$ &$\textrm{Re}(k)$ & $-\textrm{Im}(k)$ \\
      \hline
   $ 0 $&$0$& 1.40021  & 0.32376   \\
      \hline
  $ 1 $&$0$& 2.00824  &  0.340953 \\
      \hline
  $ 1 $& $1$& 1.82987  & 0.991697   \\
      \hline
  $ 2 $&$0$& 13.399173   & 15.362158    \\
     \hline
$ 2 $&$1$& 2.52447  &  1.05096 \\
    \hline
$ 2 $&$2$& 2.19558   & 1.73277 \\
\hline
\end{tabular}
\end{center}
\end{table}
\section{Final Remarks}
We studied the scalar perturbations of the full black $p$-brane solutions of ten dimensional type IIB Supergravity. The near horizon limit of extremal $p-$branes is an $AdS_{(p+2)}\times S^{(8-p)}$ space-time, which is dual to a $(p+1)$-dimensional conformal field theory at zero temperature. If we have an event horizon, the near horizon limit is a $(p+2)$-dimensional AdS black hole times a sphere $S^{(8-p)}$, dual to a field theory at finite temperture in $(p+1)$ dimensions.  We obtained the same quasinormal spectrum using the standard procedure of considering a probe scalar field in the backgroud geometry with a gauge invariant formalism. The quasinormal modes structure in such a complex problem is amazingly
simple. Allowing for a non vanishing separation constant, later
related to the glueball mass, the result is also very simple,
displaying an almost scaling behaviour. The
tensor and scalar modes are exactly the same, leading to a
simplicity of the results as well. Implications for the quark-gluon-plasma using the AdS/CFT relation awaits further analysis.

\begin{acknowledgments}
This work has been supported by FAPESP and CNPq, Brazil as well as
ICTP, Trieste.
\end{acknowledgments}

\end{document}